\documentclass[preprint,prl,nofootinbib]{revtex4}

\usepackage{graphicx}
\usepackage{dcolumn}
\usepackage{bm}

\newcommand{\beq}{\begin{equation}}
\newcommand{\eeq}{\end{equation}}
\newcommand{\beqa}{\begin{eqnarray}}
\newcommand{\eeqa}{\end{eqnarray}}

\newcommand{\cms}{{\mathrm{cm}}}
\newcommand{\second}{{\mathrm{s}}}
\newcommand{\keV}{\mathrm{keV}}
\newcommand{\GeV}{\mathrm{GeV}}
\newcommand{\TeV}{\mathrm{TeV}}
\newcommand{\MeV}{\mathrm{MeV}}

\newcommand{\Tkd}{T_{\mathrm{kd}}}

\begin{document}

\title{Early Annihilation and Diffuse Backgrounds in $1/v$ WIMP models}

\author{Marc Kamionkowski$^1$ and Stefano Profumo$^2$}
\affiliation{$^1$California Institute of Technology, Mail Code 130-33,
Pasadena, CA 91125\\
$^2$Santa Cruz Institute for Particle Physics and Department of
Physics, University of California, Santa Cruz, CA 95064}

\date{\today}

\begin{abstract}
\noindent Several recent studies have considered modifications to the
standard weakly-interacting massive particle (WIMP) scenario
in which the cross section (times relative velocity $v$) for
pair annihilation is
enhanced by a factor $1/v$.  Since $v\sim10^{-3}$ in the
Galactic halo, this may boost the annihilation rate into photons
and/or electron-positron pairs enough to explain
several puzzling Galactic radiation signals.  Here we
show that if the annihilation cross section scales as $1/v$, then
there is a burst of WIMP annihilation in the first 
dark-matter halos that form at redshifts $z\sim100-200$.  If the
annihilation is to gamma rays in the energy range
$100\,\keV-300\,\GeV$, or to electron-positron pairs in the
energy range $\GeV-2\, \TeV$, then there remains a contribution
to the diffuse extragalactic gamma-ray background today.  Upper
limits to this background provide constraints to the
annihilation cross section.  If the photon or electron-positron
energies fall outside these energy ranges, then the radiation is
absorbed by the intergalactic medium (IGM) and thus ionizes and
heats the IGM.  In this case, cosmic microwave background
constraints to the ionization history also put limits on
the annihilation cross section.
\end{abstract}


\maketitle

In the standard weakly-interacting massive particle (WIMP)
scenario \cite{Jungman:1995df,Bergstrom:2000pn,Bertone:2004pz},
dark matter is composed of particles that have
electroweak interactions with ordinary matter.  Such particles
cease to annihilate to standard-model particles in the
primordial plasma to produce a cosmological relic density that is
generically in the ballpark of that required to account for the
dark matter.  The most widely studied WIMPs are the neutralino,
the lightest supersymmetric particle in several supersymmetric
extensions to the standard model \cite{Jungman:1995df}, and, more recently, the lightest
Kaluza-Klein particle in models with universal extra
dimensions \cite{Hooper:2007qk}.  

Typical values for the WIMP mass are $m_\chi \sim 10\,\GeV -
{\mathrm{few}}~\TeV$, and freezeout of annihilations (chemical
decoupling) occurs at temperatures $T_f\sim m_\chi/20$.  After
freezeout, the WIMP temperature remains fixed to the temperature
of the primordial
plasma via frequent elastic scattering from standard-model
particles.  When the temperature drops below the
kinetic-decoupling temperature $\Tkd$, which generally falls in
WIMP models in the range $\Tkd \sim10~\MeV-\mathrm{few}\, \GeV$
\cite{Profumo:2006bv,Bringmann:2006mu}, WIMPs kinetically decouple from the
plasma, and their temperature subsequently decays with the scale
factor $R$ as $R^{-2}$, rather than $R^{-1}$.
Afterwards, WIMPs are effectively collisionless; they behave in
the subsequent Universe like the cold dark matter required to
account for detailed measurements of the cosmic microwave
background (CMB) and large-scale structure.

There still remains the possibility, though, that once galactic
halos form later in the Universe, a tiny fraction of WIMPs could
annihilate to produce cosmic gamma rays or cosmic-ray
antiprotons or positrons that might be observed
\cite{Jungman:1995df,Bertone:2004pz}.
In the standard WIMP scenario, the
annihilation cross section (times relative velocity $v$) can be
cast as
\begin{equation}\label{eq:std}
     (\sigma v)_{\mathrm{ann}}\simeq a + b (v/c)^2 +\cdots.
\end{equation}
Roughly speaking, this cross section must evaluate at $v\sim
c/2$ to $(\sigma v)_{\mathrm{ann}} \simeq
3\times10^{-26}\,\cms^3~\second^{-1}$ to obtain the correct relic
density, while annihilation in the Galactic halo occurs at
$v\sim10^{-3}$, where $(\sigma v)_{\mathrm{ann}} \simeq a
\lesssim 3\times10^{-26}\,\cms^3~\second^{-1}$.  The fluxes of
cosmic rays from WIMP annihilation are thus generically expected
to be very small; detection would require some mechanism to
boost the annihilation rate typically by a few orders of magnitude \cite{Bertone:2004pz,Baltz:2008wd}.

In recent years, there have been several unidentified Galactic
radiation backgrounds.  These include a diffuse
synchrotron haze around the Galactic center in the WMAP data
\cite{haze}; a
511-MeV gamma-ray excess from the Galactic center \cite{integral}; an excess
high-energy gamma-ray background from the Galactic center \cite{grs}; and
most recently, a reported detection of an upturn in
the cosmic-ray positron fraction at high energies \cite{pamela}.  Efforts have
been made to explain each one (or sometimes several) of these
Galactic radiation backgrounds in terms of WIMP annihilation.
In each case, though, some
mechanism to boost the annihilation rate, relative to the
natural expectation, has been introduced.

One way to boost the annihilation rate is to alter the
underlying particle theory so that the WIMP annihilation cross
section (times relative velocity) goes as $1/v$ (rather than to
a constant, as in the standard scenario) as $v\rightarrow0$.  Such a
cross section is consistent with $s$-wave unitarity
\cite{Griest:1989wd} as long as $(\sigma v)_{\mathrm{ann}} \leq
4\pi/(m_\chi^2 v)$ as $v\rightarrow0$.  Even
larger annihilation cross sections are conceivable if higher
partial waves contribute.  Specific mechanisms for providing a
low-velocity enhancement include the Sommerfeld enhancement
and/or the formation of bound particle-antiparticle states in
which the particles then annihilate
\cite{Hisano:2004ds}
although the functional dependence is not always precisely
$1/v$, an issue we elaborate on further below.

Here we parametrize the $1/v$ enhancement by writing
$(\sigma v)_{\mathrm{ann}} = 3\times10^{-26}\,
\sigma_{26}(c/v)\,\cms^3~\second^{-1}$, and we refer to these models
as ``$1/v$ WIMP models,'' or just ``$1/v$ models.''  Since the
cross section at freezeout required to obtain the relic
abundance is roughly $3\times10^{-26}\,\cms^3~\second^{-1}$, we
infer that the parameter $\sigma_{26}$ must be $\sigma_{26}
\lesssim 1$ or else the WIMP abundance will be too small.  The
equality is obtained if there are no other contributions to the
annihilation cross section; $\sigma_{26}$ could be smaller if
there are other contributions, such as those in
Eq.~(\ref{eq:std}), to the annihilation cross section.

We can make the relic-abundance constraint to $\sigma_{26}$ more
precise by including the $1/v$ dependence of $(\sigma
v)_{\mathrm{ann}}$ in the solution to the Boltzmann equation
\cite{Scherrer:1985zt}.  Doing so, we estimate the relic
abundance of $1/v$ models as
\begin{equation}
\frac{\Omega_\chi h^2}{0.1} \simeq \frac{1}{2} \sqrt{\frac{\pi}{x_f}}  
\frac{1}{\sigma_{26}}
\end{equation}
with $x_f=m_\chi/T_f \sim 20$.  This implies
$\sigma_{26}\lesssim0.2$ to explain the observed dark-matter
density, the inequality occurring, again, if there are other
contributions to the annihilation cross section.

We now describe the bounds to this scenario that arise
from upper limits to the diffuse extragalactic gamma-ray
background and from CMB constraints to the ionization history.
After the Universe becomes matter dominated, perturbations in
the cold-dark-matter density are amplified via gravitational
infall.  The smallest structures undergo gravitational collapse
first, and then more massive structures collapse later.
The kinetic coupling of WIMPs at temperatures $T_f
\gtrsim T \gtrsim T_{\mathrm{kd}}$ erases primordial structure
on mass scales smaller than \cite{Loeb:2005pm}
\begin{equation}
   M_c \simeq 33 \, (\Tkd/10\,\MeV)^{-3} \, M_\oplus.
\end{equation}
The first gravitationally-bound structures in the hierarchy
therefore have masses $M_c$.  The first objects collapse at a
redshift that can be approximated over the range $10^{-6}\,
M_\oplus \lesssim M_c \lesssim 100\,M_\oplus$ relevant for WIMPs
\cite{Profumo:2006bv} by $z_c = 140 -\log_{10}
(M_c/M_\oplus)$.\footnote{This approximation is obtained with a
scalar spectral index $n_s=1$.  It will change slightly for
$n_s\simeq0.95$.  We include the $z_c$ and $M_c$ dependences
separately in subsequent expressions so that a different
$z_c$-$M_c$ relation can be used.}
These protohalos collapse to a virial density $\rho$ that is $\approx178$
times the mean cosmological density at the collapse redshift
$z_c$.  The velocity dispersion in the protohalos can be
approximated by $v\sim R/t$, where $t\sim(G\rho)^{-1/2}$ is the
dynamical time and $R \sim (M/\rho)^{1/3}$ is the size of the halos.  Thus,
$v\sim M^{1/3} G^{1/2} \rho^{1/6}$.  Numerically, the first
halos of mass $M_c$ that collapse at redshift $z_c$ will have
velocity dispersions
\begin{equation}
     (v/c) \sim 6.0\times 10^{-9} \, (M_c/M_\oplus)^{1/3}
     (z_c/200)^{1/2}.
\label{eqn:velocity}
\end{equation}

The rate at which WIMPs annihilate in this first generation of
halos is $\Gamma = n_\chi (\sigma v)_{\mathrm{ann}}$, which
evaluates to
\begin{equation}
     \Gamma = 2.2\times10^{-17} \left(\frac{M_c}{M_\oplus} \right)^{-1/3}
     \sigma_{26} B_{2.6} \left(\frac{z_c}{200} \right)^{5/2}
     \left(\frac{m_\chi}{\TeV} \right)^{-1}\, \second^{-1}.
\label{eqn:rate}
\end{equation}
Here we have included a boost factor \cite{Kamionkowski:2008vw}, 
\begin{equation}
     B\equiv \frac{\int \rho^2 dV}{V \rho_v^2} = \frac{ c_v^3
     g(c_v)}{3[f(c_v)]^2},
\end{equation}
for a Navarro-Frenk-White density profile \cite{Navarro:1996gj},
where $c_v$ is the concentration parameter,
$g(c_v)=(1/3)[1-(1+c_v)^{-3}]$, and $f(c_v)= \ln c_v -
c_v/(1+c_v)$.  This boost factor takes into account the increase in
the annihilation rate due to the fact that the dark matter is
distributed in these first halos with an NFW density profile
$\rho(r)$, rather than uniformly distributed with density
$\rho$.  The boost factor varies from $B=2.6$ for $c_v=1$ to
$B=50$ for $c_v=10$.  To be conservative, we adopt $B=2.6$, but
include the $B$ dependence through the parameter $B_{2.6}\equiv
B/2.6$ in all subsequent expressions.  Strictly speaking, one
should also integrate $(\sigma v)_{\mathrm{ann}}$, which now
depends on $v$, over the velocity distribution in the halo.
Doing so might give an order-unity correction to the
annihilation rate, Eq.~(\ref{eqn:rate}); we fold this
theoretical uncertainty into the parameter $B_{2.6}$.

The first generation of halos survive roughly a Hubble time
before they merge into slightly higher-mass (and lower-density)
halos.  The age of the Universe at redshifts $z\gg 1$ is $t
\simeq 2\times10^{14}\, (z_c/200)^{-3/2}\, \second$.  We thus infer that the
fraction of dark-matter particles that annihilate in the first
generation of halos is
\begin{equation}
     f \simeq \Gamma t \simeq 4.4\times10^{-3} \, \left(\frac{M_c}{M_\oplus} \right)^{-1/3} \sigma_{26} B_{2.6}
     \left(\frac{z_c}{200} \right)
     \left(\frac{m_\chi}{\TeV} \right)^{-1}.
\label{eqn:fraction}
\end{equation}

Before considering bounds from diffuse radiation backgrounds, we note that a
relatively weak bound to the parameter space can be obtained
by noting that measurements of the CMB power spectrum constrain
the matter density at recombination to be within 10\% of its
value today.  We thus infer that $f\lesssim0.1$, independent of
any knowledge of the annihilation products.

However, if $1/v$ models are introduced to explain Galactic
radiation backgrounds, they require annihilation into
$e^+e^-$ pairs and/or gamma rays.  As shown in Fig.~2 of
Ref.~\cite{Chen:2003gz},
photons injected with energies $E_i$ at redshifts $z_c \sim
100-200$ in the range
$100\,\keV \lesssim E_i \lesssim300$~GeV propagate freely through the
Universe, with energies that decrease with redshift as they
propagate.  They thus appear to us as a diffuse extragalactic
background of gamma rays with energy $E_\gamma=E_i/z_c$.
Photons injected at $z\sim100$
with energies $E_i\lesssim 100$~keV propagate at first
through the Universe but then get absorbed at lower redshift by
the intergalactic medium (IGM).  Photons injected with energies
$\gtrsim300$~GeV get absorbed immediately by the IGM.

Electron-positron pairs injected into the Universe at redshifts
$z\sim100-200$ with energy $E_e$ very rapidly inverse-Compton
scatter CMB photons resulting in a gamma ray of energy $E_i
\sim (E_e/m_e)^2 T_{\mathrm{CMB}}$, where
$T_{\mathrm{CMB}}\sim10^{-2}$ eV is the characteristic
CMB-photon energy at these redshifts.  Thus, electron-positron
pairs injected with energies in the range $\GeV \lesssim E_e
\lesssim 2\, \TeV$ produce photons in the energy range $100\,
\keV \lesssim E_i \lesssim 300\, \GeV$ of the transparency
window; these gamma rays then appear to us as a diffuse
background.  Electrons injected at $z\sim100-200$ with  energies
$E_e\lesssim$GeV get absorbed by the IGM at low redshifts, and
those at $E_e \gtrsim 2$~TeV are absorbed by the IGM immediately
at high redshift.

If the photons (or electron-induced photons) are not absorbed by
the IGM, then the energy density in photons today from WIMP
annihilation in the first halos is simply $\rho_\gamma =
f\rho_\chi^0/z_c$; i.e. the fraction of the WIMP energy density
that gets converted to radiation through annihilation, scaled by
the redshift of the photons.  This evaluates numerically
to\footnote{Ref.~\cite{Oda:2005nv} also considered the gamma-ray
background from WIMP annihilation in the first objects, although
they did not consider the $1/v$ enhancement.}
\begin{equation}
     \rho_\gamma = 2.64 \times10^{-11} \,
     \left(\frac{M_c}{M_\oplus} \right)^{-1/3}
     \sigma_{26} B_{2.6} \left(\frac{m_\chi}{\TeV} \right)^{-1} \,
     \GeV~\cms^{-3}.
\end{equation}
We now compare this with the upper limit $\rho_\gamma \lesssim
5.7 \times10^{-16} \, (E_\gamma/\GeV)^{-0.1}\, \GeV~\cms^{-3}$
\cite{Sreekumar:1997un,Chen:2003gz} to obtain the constraint,
\begin{equation}
     \sigma_{26} \lesssim 2.2 \times 10^{-5} \, B_{2.6}^{-1}
     \left(\frac{M_c}{M_\oplus}
     \right)^{1/3} \left(
     \frac{E_\gamma}{\GeV} \right)^{-0.1}
     \left(\frac{m_\chi}{\TeV} \right).
\label{eqn:limit}
\end{equation}
The EGRET upper bound used here is derived for $30\, \MeV
\lesssim E_\gamma \lesssim 100\,\GeV$. It is a fit to the
extragalactic gamma-ray background, much of which comes from unresolved astrophysical
sources, and so it should be viewed as a conservative upper
limit.  Moreover, the recently launched Fermi-GLAST telescope
should soon significantly improve the upper limit to this
gamma-ray background.  Eq.~(\ref{eqn:limit}) is valid as a
conservative upper limit all the way down to energies $\sim$keV.
The true upper limits from x-ray- and gamma-ray-background
measurements in the energy range $10\,\keV \lesssim E_\gamma
\lesssim 30\,\MeV$ are more stringent.  Notice that the limit in Eq.~(\ref{eqn:limit}) does not depend on the collapse redshift $z_c$.

There will continue to be WIMP annihilation during
later, more massive, stages in the structure-formation
hierarchy.  These will produce higher-energy photons (since they
get redshifted less), but the energy density in these
higher-energy photons will be far smaller.  This can be seen
from Eq.~(\ref{eqn:fraction}) by noting that the factor $M_c$
that appears therein will be replaced by $M_*(z)$, the
characteristic halo mass at redshift $z$ (interestingly enough,
all other redshift dependence in that equation cancels).  The
mass scale $M_*(z)$ evolves very rapidly with $z$; e.g., it
goes from $10^{-12}\,M_\odot$ to $10^{-4}\, M_\odot$ from
$z\simeq200$ to $z\simeq100$, and then all the way to
$\sim10^{14}\,M_\odot$ at $z=0$.  The constraint to the model
from the diffuse-background flux is thus strongest from the
earliest halos.

Let's now consider what happens if the photon (or
electron-induced photon) falls outside the transparency window;
i.e., if it is injected with energy $E_i \lesssim 100$ keV
or $E_i\gtrsim300$ GeV.  In both cases, the photons are
absorbed by the IGM before they can reach us, and so there is no
constraint from diffuse backgrounds.  In both cases, though,
very stringent constraints arise from measurements of CMB
temperature and polarization
\cite{Chen:2003gz,Pierpaoli:2003rz,Zhang:2007zzh}.
The energy deposited in the IGM by the photons ionizes and heats
the IGM.  The reionized electrons scatter CMB photons thus
altering the observed CMB temperature/polarization power
spectra.  Ref.~\cite{Zhang:2007zzh} carried out detailed fits to
WMAP3 and large-scale-structure data to constrain the
heating/ionization of the IGM.   We infer from Fig.~2 in
Ref.~\cite{Zhang:2007zzh} that no more than a fraction
$f\lesssim 10^{-9}$ of the rest-mass energy of the dark matter
could have been injected into the IGM at a time
$\sim10^{15}$~sec after the big bang.  Although the analysis is
detailed, the magnitude of the upper limit can be understood
relatively simply: Dark matter outweighs baryons by a factor of
6, and it requires a fraction $(10\, {\mathrm{eV}}/\GeV) \sim
10^{-8}$ of the rest-mass energy of each atom to ionize it.  

Thus, if dark matter annihilates to photons with energies
$E_i\gtrsim 300$~GeV or to electron-positron pairs with
energies $E_e \gtrsim2$~TeV, then we require
\begin{equation}
     \sigma_{26} \lesssim  2.3\times 10^{-7}\,
     \left(\frac{M_c}{M_\oplus} \right)^{1/3} B_{2.6}^{-1}
     \left(\frac{z_c}{200} \right)^{-1}
     \left(\frac{m_\chi}{\TeV} \right).
\label{eqn:CMBbound}
\end{equation}
The result for photons injected with energies $E_\gamma\lesssim
100$~keV (or electron-positron pairs with $E_e\lesssim$GeV) is
similar, but weakened possibly by the redshift of the photon energy
that occurs between the time it was injected and the time it was
absorbed.  The detailed suppression depends on the injected
energy and redshift.  However, in no case is the suppression
stronger than a factor $z_c^{-1}$.  We thus conclude that the
bound will be no more than two orders of magnitude weaker than
that quoted in Eq.~(\ref{eqn:CMBbound}).

To summarize:  There will be a burst in $1/v$ WIMP models of
annihilation in the first gravitationally bound dark-matter
halos when they first form at
redshifts $z\sim100-200$.  There is a weak, albeit
final-state--independent, CMB bound that amounts to demanding
that no more than $\sim10\%$ of the dark matter annihilates
after recombination.  If the WIMP annihilates to photons in the
energy range $100\, \keV \lesssim E_i \lesssim 300$~GeV or to
electrons in the energy range GeV$\lesssim E_e \lesssim 2$~TeV,
then there are constraints, summarized in
Eq.~(\ref{eqn:limit}), to the cross section for annihilation to
$e^+e^-$ pairs and photons.  If the photons or electrons are injected
outside the transparency window, then there is a bound, quoted in 
Eq.~(\ref{eqn:CMBbound}), that comes from CMB constraints to
ionization of the IGM.

To clarify, the bounds to $\sigma_{26}$ in
Eqs.~(\ref{eqn:limit}) and (\ref{eqn:CMBbound}) are to the cross
section for annihilation to photons and $e^+e^-$ pairs, not to the
total annihilation cross section. A large hadronic final state
branching ratio is however greatly constrained by antiproton
data \cite{Bergstrom:2006tk}.  Note also that (i) $B_{2.6}$ is
likely
larger than unity, that (ii) $z_c$ is generally larger than 100 for typical WIMP
models; and that (iii) $M_c$ is generally smaller than $M_\oplus$.  The
numerical values in Eqs.~(\ref{eqn:limit}) and
(\ref{eqn:CMBbound}) will, for typical WIMP setups, generally be
smaller.  On the other hand, the numerical value in
Eq.~(\ref{eqn:limit}) may be larger for smaller $E_i$.

For the nominal WIMP values that we have chosen, the bounds are
violated by 5-6 orders of magnitude, and the constraints are
still $\sigma_{26} \ll 1$ for almost any combination of the
parameters $E_\gamma$, $M_c$, and $z_c$ consistent with typical
WIMP values.  It is thus clear that something else must be invoked
in a $1/v$ model to account for the suppression of the gamma-ray and/or
ionization-history constraints.

First consider the Sommerfeld enhancement~\cite{Hisano:2004ds}.
The $1/v$ scaling is valid only for $(v/c) \gtrsim
(m_\phi/m_\chi)$, where $m_\phi$ is the mass of a light
exchanged particle.  At smaller velocities, the $1/v$ enhancement
saturates at  $m_\chi/m_\phi$.  Our bounds can therefore
be written for this model, roughly speaking, by including a
factor $\mathrm{max}[1,(c/v)(m_\phi/m_\chi)]$, with $v/c$
evaluated from Eq.~(\ref{eqn:velocity}), on the right-hand sides
of our upper limits, Eq.~(\ref{eqn:limit}) and
(\ref{eqn:CMBbound}).  Thus, for example, for our canonical
values [$m_\chi=$~TeV, $M_c=M_\oplus$, $z_c=200$,
$B_{2.6}=1$], our limits are unaltered for $m_\phi \lesssim
6$~keV.  For larger $m_\phi$, they are reduced accordingly.
For example, the CMB bound [Eq.~(\ref{eqn:CMBbound})] is
weakened to $\sigma_{26}\lesssim 1$ (for our canonical values)
for $m_\phi\gtrsim 26$~GeV.

We mention two ways to evade the bound, even if the $1/v$
scaling extends to $v=0$, one astrophysical, and one of
particle-physics origin:  The first possibility would be to
speculate that the first microhalos that form at redshift
$z\sim100-200$ remain largely intact through subsequent stages
in the merger hierarchy.  The Galactic halo would then
consist of a huge number of distinct $\lesssim M_\oplus$
microhalos, each of which has a density $\sim 10^6$ times
the mean halo density and velocity dispersion $\sim 10^6$ times
the Milky Way velocity dispersion.  In this way, a far larger
($\sim10^{12}$) boost factor could be obtained for annihilation
in the Milky Way halo than the value $v^{-1}\sim10^3$ obtained
by assuming a smooth halo.  This would then allow a far smaller
value $\sigma_{26}$ to produce the annihilation required to
explain some of the unidentified Galactic radiation backgrounds.
Smaller values of $\sigma_{26}$ would however require the $1/v$-{\it
in}dependent terms in the annihilation cross section to
obtain the correct relic abundance.
This solution would require the survival of a significant
fraction of early microhalos; the precise survival probability
remains debated in the literature.

A second possibility is to design a model in which the cross
section for a WIMP to elastically scatter from photons and
neutrinos is also enhanced.  This would allow the WIMP to
remain in kinetic equilibrium until lower temperatures, resulting in a larger $M_c$ and a
later stage in the structure-formation hierarchy.  In turn, this would imply a suppression of the limits from the diffuse gamma-ray background (or the ionization
of the IGM) accordingly.  In the
extreme scenario, dark-matter halos would not form until
$z_v\sim10$, when the first halos and the stars they house would
have had to form to reionize the Universe.  We leave the
construction of such models for future work. 

\smallskip
\acknowledgments
We acknowledge useful comments from N.~Arkani-Hamed, J.~Feng,
D.~Finkbeiner, R.~Gilmore, M.~Pospelov, M.~Ritz, and
N.~Weiner. This work was supported at Caltech by DoE
DE-FG03-92-ER40701 and the Gordon and Betty Moore Foundation.


\begin{thebibliography}{}

\bibitem{Jungman:1995df}
  G.~Jungman, M.~Kamionkowski and K.~Griest,
  Phys.\ Rept.\  {\bf 267}, 195 (1996)
  [arXiv:hep-ph/9506380].

\bibitem{Bergstrom:2000pn}
  L.~Bergstrom,
  Rept.\ Prog.\ Phys.\  {\bf 63}, 793 (2000)
  [arXiv:hep-ph/0002126].

\bibitem{Bertone:2004pz}
  G.~Bertone, D.~Hooper and J.~Silk,
  Phys.\ Rept.\  {\bf 405}, 279 (2005)
  [arXiv:hep-ph/0404175].

\bibitem{Hooper:2007qk}
  H.~C.~Cheng, J.~L.~Feng and K.~T.~Matchev,
  Phys.\ Rev.\ Lett.\  {\bf 89} (2002) 211301
  [arXiv:hep-ph/0207125]; G.~Servant and T.~M.~P.~Tait,
 Nucl.\ Phys.\  B {\bf 650}, 391 (2003)
 [arXiv:hep-ph/0206071];
  for a review, see
  D.~Hooper and S.~Profumo,
  Phys.\ Rept.\  {\bf 453}, 29 (2007)
  [arXiv:hep-ph/0701197].

\bibitem{Profumo:2006bv}
  S.~Profumo, K.~Sigurdson and M.~Kamionkowski,
  Phys.\ Rev.\ Lett.\  {\bf 97}, 031301 (2006)
  [arXiv:astro-ph/0603373].

\bibitem{Bringmann:2006mu}
  T.~Bringmann and S.~Hofmann,
  JCAP {\bf 0407}, 016 (2007)
  [arXiv:hep-ph/0612238].

\bibitem{Baltz:2008wd}
  E.~A.~Baltz {\it et al.},
  JCAP {\bf 0807}, 013 (2008)
  [arXiv:0806.2911 [astro-ph]].

\bibitem{haze}
  D.~P.~Finkbeiner,
  Astrophys.\ J.\  {\bf 614}, 186 (2004)
  [arXiv:astro-ph/0311547];
  G.~Dobler and D.~P.~Finkbeiner,
  Astrophys.\ J.\  {\bf 680}, 1222 (2008)
  [arXiv:0712.1038 [astro-ph]];
  D.~Hooper, D.~P.~Finkbeiner and G.~Dobler,
  Phys.\ Rev.\  D {\bf 76}, 083012 (2007)
  [arXiv:0705.3655 [astro-ph]].

\bibitem{integral}
  P.~Jean {\it et al.},
  Astron.\ Astrophys.\  {\bf 407}, L55 (2003)
  [arXiv:astro-ph/0309484];
  J.~Knodlseder {\it et al.},
  Astron.\ Astrophys.\  {\bf 411}, L457 (2003)
  [arXiv:astro-ph/0309442];
  G.~Weidenspointner {\it et al.},
  arXiv:astro-ph/0406178.
  
\bibitem{grs}
  A.~W.~Strong {\it et al.},
  Astron.\ Astrophys.\  {\bf 444}, 495 (2005)
  [arXiv:astro-ph/0509290];
  D.~J.~Thompson, D.~L.~Bertsch and R.~H.~.~O'Neal,
  arXiv:astro-ph/0412376.

\bibitem{pamela}
  Talk by M. Boezio at the IDM08 conference, Stockholm and at
  ICHEP08 conference, Philadelphia; J. Chang et al. (ATIC)
  (2005), prepared for 29th International Cosmic Ray Conferences
  (ICRC 2005), Pune, India, 31 Aug 03 - 10 2005.

\bibitem{Hisano:2004ds}
  J.~Hisano, S.~Matsumoto, M.~M.~Nojiri and O.~Saito,
  Phys.\ Rev.\  D {\bf 71}, 063528 (2005)
  [arXiv:hep-ph/0412403];
  J.~March-Russell, S.~M.~West, D.~Cumberbatch and D.~Hooper,
  JHEP {\bf 0807}, 058 (2008)
  [arXiv:0801.3440 [hep-ph]];
  N.~Arkani-Hamed, D.~P.~Finkbeiner, T.~Slatyer and N.~Weiner,
  arXiv:0810.0713 [hep-ph];
  M.~Pospelov and A.~Ritz,
  arXiv:0810.1502 [hep-ph];
  J.~L.~Feng and J.~Kumar,
  arXiv:0803.4196 [hep-ph];
  J.~L.~Feng, H.~Tu and H.~B.~Yu,
  arXiv:0808.2318 [hep-ph].


\bibitem{Griest:1989wd}
  K.~Griest and M.~Kamionkowski,
  Phys.\ Rev.\ Lett.\  {\bf 64}, 615 (1990).

\bibitem{Scherrer:1985zt}
  R.~J.~Scherrer and M.~S.~Turner,
  Phys.\ Rev.\  D {\bf 33}, 1585 (1986)
  [Erratum-ibid.\  D {\bf 34}, 3263 (1986)].

\bibitem{Loeb:2005pm}
  A.~Loeb and M.~Zaldarriaga,
  Phys.\ Rev.\  D {\bf 71} (2005) 103520
  [arXiv:astro-ph/0504112];   E.~Bertschinger,
  Phys.\ Rev.\  D {\bf 74} (2006) 063509
  [arXiv:astro-ph/0607319].

\bibitem{Kamionkowski:2008vw}
  M.~Kamionkowski and S.~M.~Koushiappas,
  Phys.\ Rev.\  D {\bf 77}, 103509 (2008)
  [arXiv:0801.3269 [astro-ph]].

\bibitem{Navarro:1996gj}
  J.~F.~Navarro, C.~S.~Frenk and S.~D.~M.~White,
  Astrophys.\ J.\  {\bf 490} (1997) 493
  [arXiv:astro-ph/9611107].
  
\bibitem{Oda:2005nv}
  T.~Oda, T.~Totani and M.~Nagashima,
  Astrophys.\ J.\  {\bf 633}, L65 (2005)
  [arXiv:astro-ph/0504096].
  
\bibitem{Chen:2003gz}
  X.~L.~Chen and M.~Kamionkowski,
  Phys.\ Rev.\  D {\bf 70}, 043502 (2004)
  [arXiv:astro-ph/0310473].

\bibitem{Sreekumar:1997un}
  P.~Sreekumar {\it et al.}  [EGRET Collaboration],
  Astrophys.\ J.\  {\bf 494}, 523 (1998)
  [arXiv:astro-ph/9709257].

\bibitem{Pierpaoli:2003rz}
  E.~Pierpaoli,
  Phys.\ Rev.\ Lett.\  {\bf 92}, 031301 (2004)
  [arXiv:astro-ph/0310375];
  S.~Kasuya and M.~Kawasaki,
  Phys.\ Rev.\  D {\bf 70}, 103519 (2004)
  [arXiv:astro-ph/0409419];
  L.~Zhang, X.~L.~Chen, Y.~A.~Lei and Z.~G.~Si,
  Phys.\ Rev.\  D {\bf 74}, 103519 (2006)
  [arXiv:astro-ph/0603425];
  A.~G.~Doroshkevich, I.~P.~Naselsky, P.~D.~Naselsky and I.~D.~Novikov,
  Astrophys.\ J.\  {\bf 586}, 709 (2003)
  [arXiv:astro-ph/0208114];
  R.~Bean, A.~Melchiorri and J.~Silk,
  Phys.\ Rev.\  D {\bf 68}, 083501 (2003)
  [arXiv:astro-ph/0306357];
  S.~H.~Hansen and Z.~Haiman,
  Astrophys.\ J.\  {\bf 600}, 26 (2004)
  [arXiv:astro-ph/0305126];
  S.~Kasuya, M.~Kawasaki and N.~Sugiyama,
  Phys.\ Rev.\  D {\bf 69}, 023512 (2004)
  [arXiv:astro-ph/0309434];
  N.~Padmanabhan and D.~P.~Finkbeiner,
  Phys.\ Rev.\  D {\bf 72}, 023508 (2005)
  [arXiv:astro-ph/0503486]
  M.~Mapelli, A.~Ferrara and E.~Pierpaoli,
  Mon.\ Not.\ Roy.\ Astron.\ Soc.\  {\bf 369}, 1719 (2006)
  [arXiv:astro-ph/0603237];
  R.~Bean, A.~Melchiorri and J.~Silk,
  Phys.\ Rev.\  D {\bf 75}, 063505 (2007)
  [arXiv:astro-ph/0701224];
  S.~Kasuya and M.~Kawasaki,
  JCAP {\bf 0702}, 010 (2007)
  [arXiv:astro-ph/0608283].

\bibitem{Zhang:2007zzh}
  L.~Zhang, X.~Chen, M.~Kamionkowski, Z.~g.~Si and Z.~Zheng,
  Phys.\ Rev.\  D {\bf 76}, 061301 (2007)
  [arXiv:0704.2444 [astro-ph]].

\bibitem{Bergstrom:2006tk}
  L.~Bergstrom, J.~Edsjo, M.~Gustafsson and P.~Salati,
  JCAP {\bf 0605}, 006 (2006)
  [arXiv:astro-ph/0602632].

\end{thebibliography}
\end{document}